\documentclass[preprint,floats,aps,showpacs,epsf]{revtex4}
\usepackage{amssymb}

\usepackage{epsfig}

\newcommand{\be}{\begin{equation}}
\newcommand{\ee}{\end{equation}}
\newcommand{\bea}{\begin{eqnarray}}
\newcommand{\eea}{\end{eqnarray}}

\newcommand{\p}{\partial}
\newcommand{\s}{\sigma}

\newcommand{\la}{\langle}
\newcommand{\ra}{\rangle}
\newcommand{\rd}{\mbox{d}}
\newcommand{\ri}{\mbox{i}}
\newcommand{\re}{\mbox{e}}

\newcommand{\Tt}{{\tilde\theta}}
\newcommand{\tp}{{\theta}}

\begin{document}
\title{Low temperature correlation functions in integrable models: Derivation of the large
distance and time asymptotics from the form factor expansion}

\author{B. L. Altshuler$^*$, R. M. Konik$^\dagger$, and A. M. Tsvelik$^\dagger$}
\affiliation{$^*$ Physics Department, Princeton University, Princeton NJ 08544, 
USA;  Physics Department, Columbia University, New York, NY 10027; NEC-Laboratories America, Inc., 4 Independence Way, 
Princeton, NJ 085540,USA;\\ 
 $^\dagger$Department of  Physics, Brookhaven National Laboratory, Upton, NY 11973-5000, USA}
\date{\today}

\begin{abstract}
We propose an approach to the problem of low but 
finite temperature dynamical correlation functions in integrable 
one-dimensional models with a spectral gap. The approach is based on the analysis of the leading 
singularities of the operator matrix elements and is not model specific. We discuss only models 
with well defined asymptotic states. For such models the long time, large distance asymptotics of 
the correlation functions fall into two universality classes. These classes differ primarily by whether the 
behavior of  the two-particle S matrix at low momenta is
diagonal  or corresponds to pure reflection.  We discuss similarities and differences between our 
results and results obtained by the semi-classical method  suggested by Sachdev and Young, Phys. 
Rev. Lett. {\bf 78}, 2220 (1997).

\end{abstract}

\pacs{71.10.Pm, 72.80.Sk}
\maketitle

\sloppy
 
\section{Introduction}

In this paper we discuss the problem of finite temperature time-dependent correlation functions in 
integrable models. We do not consider here  one-dimensional models with gapless linear spectra, 
for which   there exists  a  general solution: the Matsubara $n$-point correlation functions at 
finite temperature can be  obtained  from the corresponding $T=0$ 
correlators by a conformal transformation. Instead we focus on models with spectral gaps.  
Finite temperature dynamics for that class of models is not yet fully understood. The spectral 
gap, $M$, provides an energy scale separating the high temperature ($T >> M$) and 
low temperature ($T \ll M$) regimes. 
In the former regime the leading order physics reduces to that of a conformal field theory at high temperatures.
We thus concentrate on the low temperature regime $T \ll M$.  In this limit
we will argue that the relevant physics is governed by the zero momentum limit of the interactions
between excitations in the theory.
These interactions reflect
the integrable nature of the models: the N-particle scattering matrix is a product of the 
two-particle ones. It is known that the zero momentum limit of the two-particle scattering matrix, $S(0)$, 
in an integrable system 
can be either diagonal, as it occurs, for instance, in  the 
quantum Ising model, or equal to the permutation operator. The latter alternative is realized for the 
solitons of the sine-Gordon model at general values of the coupling constant. This dichotomy suggests the 
existence of two universality classes for the low temperature dynamics. 
As we will demonstrate in this paper, this expectation is met for a class of order parameter-like operators.

The most advanced methods of evaluating correlation functions for integrable models 
with spectral gaps are based on the form factor technique (see \cite{smirnov,karowski,konik} for reviews). 
This technique allows one to 
calculate matrix elements (form factors) of the operators between the exact eigenstates 
entering the Lehmann expansion for the correlation functions. 
At finite temperature this Lehmann expansion for the two-point correlation 
function of the operator, ${\cal O}$, has the following form:
\begin{eqnarray}
G^{\cal O} (x,t) &=& \frac{1}{\cal Z} \,
{\rm Tr}(e^{-\beta H} {\cal O}(x,t) {\cal O}(0,0))\cr\cr
&=& 
\frac{\sum_{n s_n} e^{-\beta E_{s_n}}
\langle n,s_n|{\cal O}(x,t){\cal O}(0,0)|n,s_n\rangle}{\sum_{n s_n} e^{-\beta E_{s_n}}
\langle n,s_n|n,s_n\rangle}\, \cr\cr
&=&  \frac{\sum_{n,s_n;m,s_m} e^{-\beta E_{s_n}}
\langle n,s_n|{\cal O}(x,t)|m,s_m\rangle
\langle m, s_m |{\cal O}(0,0)|n,s_n\rangle}
{\sum_{n s_n} e^{-\beta E_{s_n}}
\langle n,s_n|n,s_n\rangle}\,.
\end{eqnarray}
Here the state, $|n,s_n\rangle$, denotes a set of n-particles carrying
quantum numbers, $\{s_n\}$.
The final expression in the above involves a double sum.  The first sum appears as a sum
weighted by Boltzmann coefficients while the second sum arises from 
inserting a resolution of the identity between the two fields.
The two-point correlation function has thus been reduced to a question of computing and summing
form factors.

The form factor approach turns out to be highly effective for $T=0$ two-point functions.  This is particularly so
in computing spectral functions where at finite energies the sum corresponding to that above
reduces to an expression with a finite
number of readily computable terms.
However, its generalization to multi-point functions, as well as for $T \neq 0$ faces 
difficulties caused by singularities that appear in the needed form factors \cite{smirnov}.  In this paper 
we propose a method of summation of these singularities that is suited to the extraction of the long
distance, long time, low temperature asymptotics of the correlation functions.
In a particular case, our results reproduce the semi-classical 
formulae obtained in Refs. \cite{subir,kedarsubir,zarand}.  In other instances, we find divergences between
our formulae and those derived using the semi-classical methodology proposed in Refs. \cite{subir,kedarsubir,zarand}. 


The outline of the paper is as follows.  In Section II we consider the simplest of the integrable models,
the Ising model.  Using expressions obtained in Ref. \cite{kiev} from a lattice analysis of this model, 
we analyze
the asymptotics of the spin-spin correlation functions.  In doing so we are able to cast
the spin-spin correlation function in a form nearly identical to what would be found using the Lehmann
expansion.  We take this similarity as instructive: it both indicates how to handle singularities that appear
in the Lehmann expansion and it tells us how to proceed with more general integrable models.  
Thus in the next section, Section III, we generalize our computations to arbitrary integrable models.  We find 
two universality classes of behavior.  One class of behavior is reminiscent of the Ising model.
Here we find no discrepancies between our approach and that of the semi-classical approach.  For the
second universality class we find novel scaling behavior that is not reproduced semi-classically.
We consider the implications of these two universality classes for the physics of the sine-Gordon model
in Section IV. We conclude our paper with a brief discussion of the results. 

\section{Finite Temperature Spin-Spin Correlation Functions in the Ising Model}
 
To understand how to deal with the singularities that appear in the Lehmann expansion, we
first turn to the quantum Ising model.
In the following section we will extend our approach to other integrable models.  

We first review the basics of the model.
The Hamiltonian, $H$, of the 
Ising model is equivalent to a Hamiltonian of noninteracting massive fermions
\bea
&& H = \sum_n[- J\s^z_n\s^z_{n+1} + h\s^x_n] \equiv \sum_n[- h\mu^z_{n-1/2}\mu^z_{n+1/2} + J\mu^x_n] 
\equiv \sum_p \epsilon(p)F^+_pF_p ;\nonumber\\
&& \epsilon(p) = \sqrt{(J- h)^2 + 4Jh\sin^2(p/2)}, \label{is}
\eea
where $\mu^z_{n+1/2} = \prod_{j<n}\s_j^x$ is the disorder operator and $F,F^+$ are fermion 
annihilation and creation operators. The fermions represent walls between domains 
with different orientations of the magnetization and so are to be thought of as solitons.  

Calculations simplify  in the continuum limit,  when the spectral gap $M = |J- h|$ is much smaller 
than the bandwidth $\sim J$ and the spectrum is relativistic $\epsilon(p) = \sqrt{c^2p^2 + M^2}$, $c^2 = Jh$. 
In this case, energy and momentum of a quasi-particle are conveniently parameterized by 
a rapidity, $\theta$, ($cp = M\sinh\theta$). Then the eigenstates of Hamiltonian (\ref{is}) are 
labeled  by sets of 
rapidities, $\{\theta_i\}$, such that the energy and momentum of the system are equal to
\bea
E = M\sum_{i=1}^n\cosh\theta_i, ~~ P = c^{-1}M\sum_{i=1}^n\sinh\theta_i.
\eea
Below we set $c=1$. Operators $\s^z$ and $\mu^z$ (we will call them $\s$ and $\mu$) have 
infinitely many matrix elements.  By Lorentz invariance, 
the form factors depend on rapidity differences \cite{mccoy}:
\bea
\la \theta_1,...\theta_n|\s|\theta'_1, ...\theta_m'\ra = AM^{1/8}\prod_{i<j}\tanh(\theta_{ij}/2)
\prod_{p< q}\tanh(\theta'_{pq}/2)\prod_{i,p}\coth[(\theta_i - \theta'_p \pm i\epsilon)/2], \label{form}
\eea
where at $h < J$ (the ordered phase) $n + m$ is even for $\s$ and odd for $\mu$ (for $h > J$ 
it is the other way around). Here $\theta_{ij} = \theta_i - \theta_j$ and 
$A$ is a known numerical constant. The singularities in (\ref{form})  appear as soon as  some 
$in$ and $out$ rapidities coincide. 
Singularities of this kind are known as  kinematical poles and 
appear routinely in integrable models. 
There are operators however where the matrix elements are free of singularities. 
In the Ising model such an operator is the energy 
density operator, 
\bea
\s^x(x) = \sum_k\re^{-\ri q x}\gamma(k)\gamma(k-q)\hat F_k\hat F_{q -k}, ~~ \hat F_k = 
\hat F^+_{-k}, ~~\gamma(k) = \sqrt{1 + k/\epsilon(k)}.\label{edensity}
\eea
We will not in general consider the correlation functions of such operators in the paper (although
see Section IVc).  We can say however their long time, large distance behavior will generally be ballistic.

\subsection{Possible Approaches}

In the Ising model the form factors are known explicitly and 
it has been a common belief 
that their analysis would lead to a better understanding of the general problem. 
Several approaches to the problem of the singularities that appear in the Lehmann expansion are possible:
\begin{itemize}
\item
One can try to regularize the singularities by considering a finite system. 
Understanding the singularities as arising from an infinite sized
system has been advocated both in Ref. \cite{balog} and in Ref. \cite{saleur}.
In the Ising case it is clear how this regulation would proceed.  In the infinite
system the singularities arise when the momenta of 
the $in$ and $out$ states coincide.  However if the system is treated as on a ring of finite size,
the $in$ and $out$ states must be quantized differently.  The spin operator in the Ising model
connects the Ramond and Neveu Schwarz sectors of the Hilbert space.  In the Ramond sector,
the fermionic modings are periodic while in the Neveu Schwarz sector they are anti-periodic.
(These features of the Ising model Hilbert space are reviewed in Ref. \cite{fonseca}.)
Thus at finite system size the momenta simply cannot coincide and the expressions are singularity free.  
However the summation of form factors
is then discrete not continuous.  To the best of our
knowledge no one has then succeeded in explicitly performing the necessary double summation.
\item
In a similar spirited approach, one can
interchange imaginary time and space axes using a Wick rotation. 
The finite $T$ Matsubara Green's functions of the infinite system then become 
zero temperature correlators for a system on a circle of circumference, $1/T$. 
While the corresponding form factor expansion is manifestly free of singularities, 
one faces the problem of calculating form factors and excitation energies in a system of finite size. 
For the Ising model this was achieved in Ref. \cite{kiev}.
The two-point correlation  functions of $\s$'s and $\mu$'s 
were obtained rigorously in the form of series.  
Below we use these series to extract the long time and 
distance asymptotics of the dynamical correlation functions.
\item
One can calculate correlations by re-expressing them as the determinant of
an integral operator of the Fredholm type.
In this way results were obtained for the spin-1/2 XX magnet \cite{its} and the 
Bose gas \cite{korepin}.  For a general exposition on this methodology see Ref. \cite{book}.
\item
Exploiting determinantal expressions of the matrix elements arising from explicit expressions for the Bethe
eigenfunctions, Refs. \cite{maillet,caux} have obtain expressions for the zero temperature correlation functions
in the Heisenberg XXZ chain.  It is likely this approach can be readily extended to finite temperature
computations.  
\item
In an approach adopted in Ref. \cite{saleur} one can imagine treating the computation of correlation functions at 
finite temperature in terms of a renormalized `thermal' vacuum and a renormalized set of `thermal' excitations.
The computation of correlation functions then proceeds in a fashion similar to that of the zero temperature
calculation but with the form factors obeying a revised set of axioms.  More recently, the notion of a thermal
vacuum was explored in the work of Ref. \cite{doyon}.
\item
The sole method that does not trade on the underlying integrability of the models of concern
was suggested by Sachdev and Young \cite{subir}
in the form of a ``renormalized classical'' approach valid at finite $T$ and long time.  
The result for the causal correlation function 
of the spin field in the ordered phase valid for $T \ll M$ is
\bea
D(x,t) \equiv \la \s(t,x)\s(0,0)\ra = CM^{1/4}\exp\left[- \int \frac{\rd p}{\pi}
\re^{-\epsilon(p)\beta}|x - t\frac{\p\epsilon(p)}{\p p}|\right]\label{sachdev},
\eea
where $\epsilon = \sqrt{M^2 + p^2} \approx M + p^2/2M$ and $C$ is a numerical constant. 
This expression suggests that the typical correlation length scales as $t$. One can 
approximate (\ref{sachdev}) as   $D(x,t) \approx  C\exp[ - n(T)\mbox{max}(|x|, v|t|)]$ 
where $v = \sqrt{\pi T/2M}$ is the thermal velocity of kinks and $n(T)= \sqrt{TM/2\pi}\re^{-M/T}$ 
is the average number of solitons (fermions).  
For the case of the Ising model, we show below that Eq. (\ref{sachdev}) 
also follows directly from the form factor approach using the 
results of Bugrij \cite{kiev}. 
\item
The final possible approach involves attacking directly the double sum Lehmann expansion (Eqn. 1) through
the use of infinite volume form factors.  To understand how this is to be done in general,
we employ the series expansions from Ref. \cite{kiev} which will provide the necessary clues
to understanding the singularity structure in this case.
\end{itemize}

\subsection{Long space and time asymptotics of the spin-spin correlation function}

In the section we use the results of Ref. \cite{kiev} to understand the asymptotics
of the spin-spin correlation function in the Ising model.  From this we will be able to 
follow a natural path to generalizing these results to integrable models at large.

The result for the Matsubara Green's function of the Ising model order parameter field
obtained in Ref. \cite{kiev} is 
\bea
&& \la\s(\tau,x)\s(0,0)\ra = \label{kiev1}\\
&& CM^{1/4}\re^{-2|x|n(T)}\left\{ 1 + \sum_{N=1}^{\infty}\frac{T^{2N}}{(2N)!}
\sum_{q_1,...q_{2N}}\prod_{i=1}^{2N}
\frac{\re^{-|x|\epsilon_i - \ri \tau q_i - \eta(q_i)}}{\epsilon_i}\prod_{i > j}
\left(\frac{q_i - q_j}{\epsilon_i + \epsilon_j}\right)^2\right\}, \nonumber
\eea
where $\tau$ is imaginary time, 
\bea
n(T) = \int\frac{\rd p}{2\pi}\re^{-\beta\epsilon(p)} \approx \sqrt{TM/2\pi}\re^{-M/T},
\eea
is the soliton density, $q =2\pi Tm$ ($m$ integer), and $\epsilon(q) = \sqrt{M^2 + q^2}$. The term
in the exponent,
\bea
\eta(q) = 
\frac{2\epsilon(q)}{\pi}\int_0^{\infty}\frac{\rd x}{\epsilon^2(q) + x^2}\ln\coth[\beta\epsilon(x)/2],
\eea
at $T \ll M$ is exponentially small and will be neglected from hereon.
The symmetry breaking transition at $T=0$ leads to a finite magnetization,  
$\la\s\ra = \pm [CM^{1/4}]^{1/2}$. This is reflected in the zeroth order term in Eq. (\ref{kiev1}). 
We see that this expression for the correlator has the form of a zero-temperature
Lehmann expansion: the sum over N represents intermediate states with N particles while
the sum over the $q_i$'s are momentum sums for a given set of N particles.
The rightmost term in Eq. (\ref{kiev1}) can then be interpreted as a finite volume form factor squared.
We can see that the correlation function, Eq. (\ref{kiev1}), is periodic in $\tau$, as it must be.  
The entire expression can then be thought of
as a Wick rotation of a computation done in the picture with periodic boundary conditions in 
the space direction.

The correlation function of the disorder parameters can be presented in a similar fashion,
\bea
&&\la\mu(\tau,x)\mu(0,0)\ra = \label{kiev2}\\
&&CM^{1/4}\re^{-2|x|n(T)}\sum_{N=0}^{\infty}\frac{T^{2N +1}}{(2N +1)!}\sum_{q_1,...q_{2N+1}}
\prod_{i=1}^{2N+1}\frac{\re^{-|x|\epsilon_i - \ri \tau q_i - \eta(q_i)}}{\epsilon_i}\prod_{i > j}
\left(\frac{q_i - q_j}{\epsilon_i + \epsilon_j}\right)^2 \nonumber,
\eea
where now we have a sum over an odd number of $q$'s.

\begin{figure}
\begin{center}
\epsfxsize=0.6\textwidth
\epsfbox{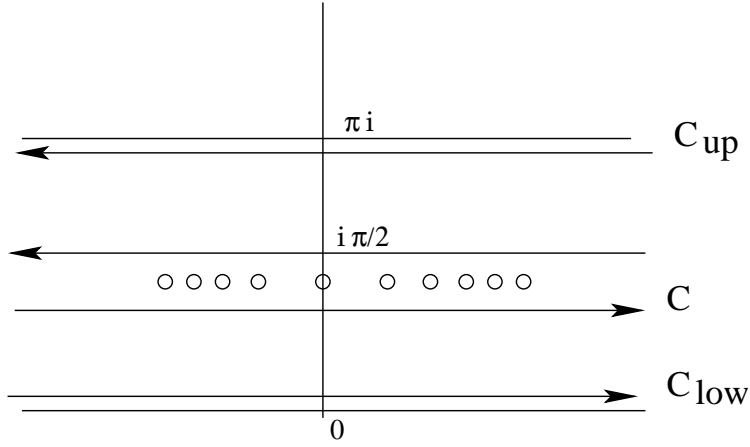}
\end{center}
\caption{The integration contours in the $\theta$ plane. The circles are poles at $\theta = \ri \pi/2 + 
\sinh^{-1}(2\pi n T/M)$. }
\label{cont}
\end{figure}

Eqs. (\ref{kiev1}) and (\ref{kiev2}) are well placed to rewrite 
the correlation functions as an 
expansion in powers of $\exp[-\beta M]$. 
Focusing on the order-parameter correlation function, we rewrite Eq. (\ref{kiev1})
representing the $N$-th terms in Eq. (\ref{kiev1}) as a sequence of contour integrals:
\bea
\frac{1}{N!}\int_{C}\rd\theta_1 \frac{\re^{\tau\cosh\theta_1 + 
\ri x\sinh\theta_1}}{\re^{\beta\cosh\theta_1} - 1}...\int_{C}\rd\theta_N 
\frac{\re^{\tau\cosh\theta_N + \ri x\sinh\theta_N}}{\re^{\beta\cosh\theta_1} - 1}
\prod_{i> j}\tanh^2[(\theta_i - \theta_j)/2],
\eea
with the contour $C$ shown  on Fig. \ref{cont}. 
One can deform this contour into the superposition $C \rightarrow C_{low} + C_{up}$, 
where $C_{low}$ lies just above real axis while $C_{up}$ corresponds to $\Im m\theta = \pi - 0$.  
As a result one obtains 
a sum of particle (lower contours) and anti-particle (upper contours) contributions:
\bea
&& \sum_{n=1}^N\frac{1}{n!(N-n)!}\int\prod_{i=1}^n\rd\theta_i f(\epsilon_i)\re^{\tau\epsilon_i + 
\ri x p_i}\prod_{j=1}^{N-n}\rd\theta'_j f(\epsilon_j)\re^{(\beta -\tau)\epsilon_j - \ri x p_j}\times\nonumber\\
&&\prod_{i>k}\tanh^2[(\theta_i - \theta_k)/2]\prod_{j> p}\tanh^2[(\theta'_i - \theta'_p)/2]\times\prod_{i,p}
\coth^2[(\theta_i - \theta'_p + \ri 0)/2] ,\label{truth}
\eea
where $f(\epsilon) = [\re^{\beta\epsilon} -1]^{-1}$ is the 
Bose distribution function and $p = M\sinh\theta, \epsilon = M\cosh\theta$. In Eq.(\ref{truth}), 
$n$ and $N-n$ are numbers of particles and antiparticles. 
Note that the singularities of the form factors are now removed from the 
real axis to one side of the contour, as was suggested in Ref. \cite{balog}. 
A formula similar to  Eq. (\ref{truth}) was proposed in 
Ref. \cite{mussardo}. However, there is  a noticeable difference:  
the distribution functions turn out to be  of the Bose type, 
although  the quasi-particles of the Ising model are fermions.

To obtain the casual Green's functions, one has to replace $\tau$ with $\ri t$.  
From now on we will be interested in the behavior 
of the Green's functions at longest real times $t \gg M^{-1}$ and $|x| \gg M^{-1}$. 
Only those terms in Eq. (\ref{truth}) which contain both particles and antiparticles are singular. 
The leading contribution to the Green's function 
in each order comes from the vicinity of each  singularity. 
Each $\theta_{in}$ carries the thermal exponent $\re^{-\beta M} \ll 1$. 
It is thus sufficient to restrict the summation in Eq. (\ref{truth}) by terms with equal numbers 
of $in$ and $out$ rapidities, each $in$ rapidity $\theta_{i}$ being 
close to one of the $out$ rapidities   $\theta'_{j}$.  There are
$n!$ ways to choose a pair for each $\theta_i$; 
this factor cancels one of the $n!$ terms 
from the denominator of Eq. (\ref{truth}).  
The integral over the difference $\theta_i - \theta_j'$ can be evaluated with the
resulting series summing up to the exponent,
\bea
\la\s(x,t)\s(0,0)\ra  = CM^{1/4}\theta(t)\re^{-2|x|n(T)}\exp[R(t,x)], \label{result}\eea
where  
\begin{eqnarray}
R(t,x > 0) &=& \frac{1}{\pi^2}\int \rd\theta_1\rd\theta_2 
\re^{-\beta M(1 + \theta_1^2/2)}\frac{\re^{\ri Mt(\theta_1^2 - \theta_2^2)/2 + \ri Mx\theta_{12}}}
{(\theta_{12} + \ri 0)^2} \nonumber\cr\cr
&=&  \frac{M\re^{-\beta M}}{\pi}\int \rd\theta \re^{-\beta M\theta^2/2}\left[|x| - |x - \theta t|\right].
\label{R}
\end{eqnarray}
This result reproduces the semi-classical computation found in Eq. (\ref{sachdev}). 

We now turn to how this result determines how we should proceed in the case of general
integrable models.

\section{The general case: Two universality classes}

The form of Eq. (\ref{truth}) has a form intimately related to the double summed Lehmann expansion
of Eq. 1.  Apart the bosonic distribution function $f(\epsilon)$ (unimportant in the limit $T \ll M$), 
Eq. (\ref{truth}) looks for all intent purpose as those subset of terms in Eq. 1 involving 
form factors (as given by Eq. 4) of exactly
N particles or excitations.  This then 
implies that the double summed Lehmann expansion is tractable.
Or put another way, we can generalize to correlation functions involving
an operator ${\cal O}$ in other integrable models 
by
using Eq. (\ref{truth}) with the Ising model form factors replaced  
by the form factors of ${\cal O}$. 
At temperatures much smaller than the gap, we replace the distribution 
functions by Boltzmann exponents. 
The asymptotics of the correlation function  is then determined solely 
by the behavior of the form factors in the vicinity of the kinematic poles. 
As we have already pointed out, 
the existence of kinematic poles is a general consequence of the theory of form factors \cite{smirnov} 
rather than a specific feature of the Ising model.  These poles are due to annihilation
processes and occur when rapidities of a particle and an anti-particle coincide.  The behavior of 
form factors in the vicinity of the poles divide all integrable models into two universality classes. 
Accordingly the resolution of the problem of singularities should not be model-specific. 

Consider a generic matrix element of operator ${\cal O}$:
\bea
 F^{\cal O}(\theta_i,a| \Tt_i\,b) 
= \la \tp_1,\ldots,\tp_n|{\cal O}|\Tt_1,\ldots,\Tt_m\ra_{ a_1,... a_n; b_1,...b_m}, \label{FF}
\eea
where $\tp_i$ and $\Tt_j$ are rapidities and $a_i$ and $b_j$ are isotopic indices of $in$ and $out$ 
particles. One can express  
the form factor (\ref{FF}) in terms of 
canonical matrix elements. Let $F^{\cal O}_{irr}$ be the irreducible part of $F^{\cal O}$ 
given by the connected diagrams. Only this part is necessary for our calculations done to 
lowest order in $\exp(- M\beta)$. 
According to Refs. \cite{balog,mussardo,konik1}, 
$F^{\cal O}_{irr}$ is equal to the matrix element 
between the vacuum and a certain excited state of $n+m$ particles:  
\bea
\label{poles}
 F^{\cal O}(\tp_i,a| \Tt_i, b)_{irr} = \la 0|{\cal O}|\Tt_1,\ldots,\Tt_m;
-\ri\pi + \tp_n \pm \ri 0,\cdots , -\ri\pi + \tp_1 \pm \ri 0 \ra_{b_1,...b_m; \bar a_n,...\bar a_1;}
\eea
where $\bar a$ indices are obtained from $a$ ones by charge conjugation.  
The choice of whether the small imaginary part, $\ri 0$, is positive or negative
determines the exact form of the disconnected parts (here ignored) of the form factor in Eq. (\ref{FF}).

\begin{figure}
\begin{center}
\epsfxsize=0.6\textwidth
\epsfbox{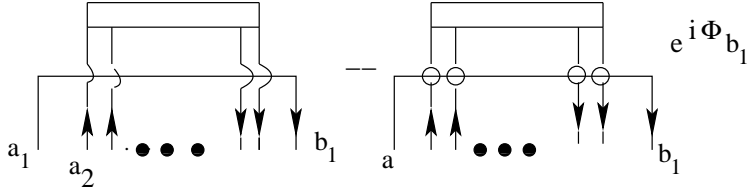}
\end{center}
\caption{A graphical representation of Eq. (\ref{gen}) 
written in terms of the irreducible form factors in Eq. (\ref{FF}) (the rectangles). 
The up arrows represent rapidities $\Tt_i$, while
the down ones correspond to the rapidities $-\ri\pi + \tp_j$. 
Circles at the intersections of the solid lines represent two-particle $S$ matrices.}
\label{ann}
\end{figure}

The axiom for the annihilation poles in the standard formulation is expressed in terms of the 
form factors of the type
\begin{equation}
{\cal F}^{\cal O}_{a_1,\ldots,a_n}(\theta_1,\ldots,\theta_n) 
= \la 0|{\cal O}|\theta_n,...\theta_1\ra_{a_n,\ldots,a_1}.
\end{equation}
Namely, as was shown in Ref. \cite{smirnov} 
(see also Ref. \cite{konik} in review), 
these form factors have poles at $\theta_n - \theta_j = \ri\pi$, $j<n$, 
and the residues satisfy the following relation (see Fig. \ref{ann}):
\bea
&&\ri {\rm Res}_{\theta_{n} = \theta_{n-1}+\ri\pi} 
{\cal F}^{\cal O} _{a_1,\ldots,a_n}(\theta_1,\theta_2,\ldots,\theta_n) =   
{\cal F}^{\cal O} _{a'_1,\ldots,a'_{n-2}}(\theta_1,\ldots,\theta_{n-2}) \label{gen} \\
&& \times C_{a_n,a'_{n-1}}\left[{\bf \rm I}-\re^{2\pi \ri \varphi_{a_n,{\cal O}}} 
S_{\tau_1,a_1}^{a'_{n-1},a'_1}(\theta_{n-1}-\theta_1)
\bigg( \prod_{i=2}^{n-3} S_{\tau_i,a_i}^{\tau_{i-1},a'_i}(\theta_{n-1}-\theta_i) \bigg)
S_{a_{n-1},a_{n-2}}^{\tau_{n-3},a'_{n-2}}(\theta_{n-1}-\theta_{n-2})\right] . \nonumber 
\eea 
Here $S$ is the  two-particle scattering matrix, $C$ is the charge conjugation matrix, and $\varphi_{a,{\cal O}}$
is the {\it semi-locality index} between the particle creation operator $A_a^\dagger$ and ${\cal O}$. 
We will define this index  later.

To demonstrate how Eq. (\ref{gen}) can be used to extract the most singular behavior of the form factors, 
we combine it with (\ref{FF}) to obtain the following 
behavior of two- and three-particle form factors at the poles:
\bea
\la \theta, a |{\cal O}|\theta_1,b \ra &=& -\frac{
\ri(1 - \re^{2\pi\ri\varphi_{b,{\cal O}}})}{\theta_1 - \theta \pm \ri 0}\la 0|{\cal O}|0\ra \delta_{a,b} 
+ \mbox{disconnected part;}\label{two}\\
\la \theta_1, a_1;\theta_2,a_2 |{\cal O}|\theta,a \ra &=& 
-\ri\frac{\delta_{a,a_2}\delta_{a_1,a'_1} - e^{i2\pi\varphi_{a,{\cal O}}}
\delta_{a,a'_2} S^{\bar{a_2'},\bar{a_1'}}_{\bar{a_2},\bar{a_1}}(\tp_2-\tp_1)}
{\theta-\theta_2 \pm \ri 0}\la 0|{\cal O}|\tp_1-\ri\pi,\bar{a_1'}\ra \nonumber \\ 
&& -\ri\frac{\delta_{a,a'_1}S^{\bar{a_2'},\bar{a_1'}}_{\bar{a_2},\bar{a_1}}(\tp_2-\tp_1)
- e^{i2\pi\varphi_{a,{\cal O}}}\delta_{a,a_1}\delta_{a_2,a'_2}}
{\theta-\theta_1 \pm \ri 0}\la 0|{\cal O}|\tp_2-\ri\pi,\bar{a_2'}\ra \nonumber \\ 
&& + \mbox{~~disconnected part.} 
\eea
Now it is appropriate to  define  the index $\varphi$.  
We will call two operators, ${\cal O}_1$ and ${\cal O}_2$, mutually non-local if the Euclidean
correlator $\langle \ldots {\cal O}_1({\bf r}) {\cal O}_2 (0) \ldots
\rangle$ is a multi-valued function of ${\bf r}$. 
In particular, if  
under analytic continuation $z\to z \re^{2\pi \ri}$,  
$\bar z\to \bar z \re^{-2\pi \ri}$ ($z=\tau +\ri ~x$ and $\bar z=\tau -\ri ~x$), the correlator
acquires only a phase, $2\pi \varphi_{1,2}$, the
two operators are called {\em semi-local}. 
The remarkable advantage  of this 
definition is that the index $\varphi_{1,2}$ can be evaluated at small distances, 
where all correlation functions have a simple power law form.  

From Eq. (\ref{two}) it follows that an operator ${\cal O}$ with a finite vacuum expectation
has annihilation poles only if $\varphi_{a,{\cal O}}$ is not an integer, that is when  ${\cal O}$ is non-local. 
For instance,  $\s$ and $\mu$ in the Ising model are non-local with respect to the fermions 
(they acquire a phase, $\pi$, after circling around the fermions in the $\tau-x$ plane). 
However, if the operator can create an  odd number of particles, 
the annihilation poles may appear provided the $S$ matrix is non-trivial. 
For instance, the vector field ${\bf n}$ in the O(3) nonlinear sigma model belongs to the later category. 
Its form factors do have annihilation poles, although its vacuum expectation value is zero \cite{smirnov}.  

We proceed with the calculation of  the most singular parts of the form factors. We will do
it explicitly for operators with $\la 0|{\cal O}|0\ra \neq 0$. 
The maximal singularities come from the matrix elements 
which contain equal numbers of $in$ and $out$ particles. 
The contributions from all other matrix elements contain additional powers 
of either $\exp[-\beta M]$ or $\exp[- \ri tM]$. 
Since form factors with different order of rapidities are related to one 
another (Refs. \cite{smirnov,karowski,konik}), it is sufficient
to know the residues for one particular arrangement. 
For the form factors of the type in Eq. (\ref{FF}), 
we focus on the residues at  the poles at $\theta_i = \Tt_i$. 
In this case one can use the crossing symmetry of the S-matrix and obtain from Eq. (\ref{gen}) 
the relation for the residues of (\ref{FF}).  This allows the convenient graphical representation of 
Fig. \ref{ann}.  

At this point, the computation divides itself into two cases.  Each case marks out a separate universality class.
We deal with each in turn.

\subsection{Ising-like Universality Class}

In the first case, Ising-like behavior results.  
This occurs if either of following conditions hold
\vskip 10pt
1.  The $S$ matrix is diagonal.
\vskip 10pt
2.  The $S$ matrix is not diagonal, but the semi-locality index,
$\varphi_{a,{\cal O}}$ = 1/2, $a=1,\ldots, \alpha$, is independent of the particle index, $a$. 
\vskip 10pt

\noindent In these cases the residue is diagonal in the isotopic indices: $a_i = b_i$. 
For the diagonal S-matrix this is obvious. 
For the case $\varphi = 1/2$ with a non-diagonal S-matrix, some additional reasoning is required. 
Namely, one needs to apply  Eq. (\ref{gen}) in successive steps. 
For the first step we obtain the residue in the two-particle form factor. 
This yields Eq. (\ref{two}) diagonal in the isotopic indices. 
Then we use this result to calculate the coefficient at  $[(\tp_1 - \Tt_1)(\tp_2 - \Tt_2)]^{-1}$ 
in the four-particle form factor 
${\cal F}_{\cal O}(\tp_1-\ri\pi\pm\ri 0,\tp_2-\ri\pi\pm\ri 0,\Tt_2,\Tt_1)_{\bar{a_2}\bar{a_1}b_2b_1}$. 
As follows from Eq. (\ref{gen}), the S-matrices enter into the expression for this coefficient 
in the vicinity of the double pole in the combination 
\bea
S_{\bar{a}_2,\bar{a}_1}^{\bar{a}'_2,\bar{a}'_1}(\tp_{21})S_{b_2,\bar{a}'_1}^{a'_2,\bar{a}''_1}(\ri\pi + \tp_{21}) 
= S^{\bar{a}'_2,\bar{a}'_1}_{\bar{a}_2,\bar{a}_1}(\tp_{21})S^{\bar{b}_2,\bar{b}_1}_{\bar{a}'_2,\bar{a}'_1}(-\tp_{21}) 
= \delta_{a_1,b_1}\delta_{a_2,b_2}.
\eea
Here we have used the crossing symmetry and unitarity of the S-matrices. 
As we see, this residue is again diagonal in the isotopic indices and 
equal to $(-\ri)^2(1 - \re^{2\pi\ri\varphi})^2 = -4$.

Thus for these two cases, it is clear that a multiple application of formula (\ref{gen}) 
yields the main singularity for general $n$ as
\begin{eqnarray}
F^{\cal O}_{sing}(\tp_1,\ldots,\tp_n;\Tt_1,\ldots,\Tt_n) 
&=& (-\ri)^n\sum_{P_i}(1 - \re^{2\pi \ri \varphi_{a,{\cal O}}})^n
\prod_{i=1}^n(\tp_i - \Tt_{P_i} \pm \ri 0)^{-1}\la 0|{\cal O}|0 \ra;\cr\cr
&=& (-2\ri)^n\sum_{P_i}
\prod_{i=1}^n(\tp_i - \Tt_{P_i} \pm \ri 0)^{-1}\la 0|{\cal O}|0 \ra,
\end{eqnarray}
where the sum includes all permutations of $\{1,...n\}$. 
The result for the correlation function of operators, ${\cal O}$,
with {\it finite vacuum expectation values} is then essentially the same as Eq. (\ref{sachdev}) but with 
\bea
\exp[-\beta\epsilon] \rightarrow \sum_{j=1}^\alpha\sin^2(\pi\varphi_j)\exp[-\beta\epsilon_j]
= \sum_{j=1}^\alpha\exp[-\beta\epsilon_j], \label{corr}
\eea
where the sum runs over isotopic indices of the excitations.

\subsection{Non-Ising Like Universality Class}
When both the S-matrix is non-diagonal and the semi-locality index, $\phi_{a,{\cal O}}$,
is not simply $\pm 1/2$, the behavior of the correlators is considerably more complex.
Attendantly, the calculations are more difficult 
due to the fact that $\varphi_a$ depends on the particle index.  
Nevertheless the maximal residue can be calculated, although only at small rapidities 
(in the given context, that is all we need). 
We take advantage of the fact that in integrable models with a non-diagonal $S$-matrix, 
the $S$ matrix has the universal limit $S(\theta \rightarrow 0) = -P$ 
(the permutation operator). This limit  corresponds to a pure reflection. 
In general the crossover to pure reflectivity occurs at some model-dependent scale $\theta_0$. 
It may happen that this  scale is much smaller than one; then universal dynamics occurs only  
for $T \ll M\theta_0^2 < M$ when rapidities $|\tp | \ll \theta_0$ 
determine the expressions for the correlation functions. 
At these temperatures, 
one can replace all S-matrices by permutation operators which leads to 
simplifications in Eq. (\ref{gen}) (see Fig. \ref{ann1}): 

\begin{figure}
\begin{center}
\epsfxsize=0.6\textwidth
\epsfbox{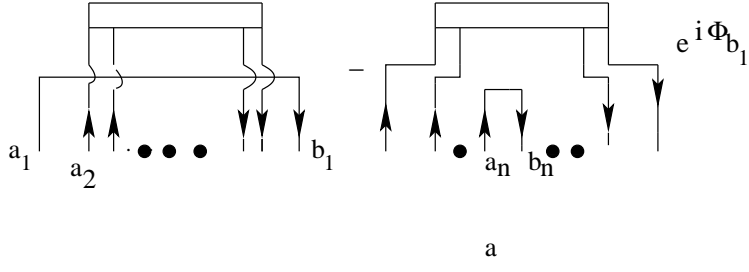}
\end{center}
\caption{A graphical representation of Eq. (\ref{gen}) for S=P.}
\label{ann1}
\end{figure}

As we see from Fig. \ref{ann1}, the isotopic indices of $in$ and $out$ particles must coincide. 
This allows us to simplify the notational representation of the form factor:
\[
\mbox{Res}_{\tp_i = \Tt_i} 
{\cal F}^{irr}(\tp_1,\ldots,\tp_n;a_1,...a_n|\Tt_n, ... \Tt_1; a_n,...a_1) 
\equiv (-i)^n f(a_1,...a_n)\la 0|{\cal O}|0\ra .\]
Then the equation depicted in Fig. \ref{ann1} becomes 
\bea 
f(a_1,...a_n) = f(a_2,...a_n) - \re^{\ri\phi_{a_n}}f(a_1,...a_{n-1}) \label{id}.
\eea
This recurrence relation can be solved explicitly with the result
\bea
f(a_1,...a_n) = \sum_{k=1}^n
\left[(1 - \re^{\ri\phi_k})(-1)^{(n-k)}\exp\left(\ri\sum_{j= k+1}^n\phi_j\right)\left(
\begin{array}{c}
n-1\\
k-1
\end{array}
\right)\right]\label{solution},
\eea
where $\phi_a \equiv 2\pi\varphi_{a,{\cal O}}$. 

In this case the correlator, $\la {\cal O}(x,t){\cal O}(0,0)\ra$, takes the
form
\begin{eqnarray}
\la {\cal O}(x,t){\cal O}(0,0)\ra  &=& \la {\cal O}\ra^2 \sum^\infty_{n=0} \frac{1}{n!}\sum_{a_i}|f(a_1,\ldots,a_n)|^2
\int \frac{d\tp_1}{2\pi}\cdots\frac{d\tp_n}{2\pi}\frac{d\Tt_1}{2\pi}\cdots\frac{d\Tt_n}{2\pi} \cr\cr
&& \hskip -1.5in \times
\exp\bigg[-\beta M\sum^n_{j=1}\cosh(\Tt_j)\bigg] \cr\cr
&& \hskip -1.5in \times \exp\bigg[iMt\sum^n_{j=1}(\cosh(\tp_j)-\cosh(\Tt_j))-iMx\sum^n_{j=1}(\sinh(\tp_j)-\sinh(\Tt_j))\bigg] \cr\cr
&& \hskip -1.5in \times \frac{1}{2^n}\sum^n_{k=0} C_n^k \prod_{q=1}^k\frac{1}{(\tp_q-\Tt_q-\ri 0)^2}\prod_{p=k+1}^{n}\frac{1}{(\tp_p-\Tt_p+\ri 0)^2},
\label{double}
\end{eqnarray}
where $C^k_n = ({n\atop k})$.  Again this expression arises either from using the form factors
of ${\cal O}$ in conjunction with Eq. (\ref{truth}) or directly from the double summed
Lehmann expansion (Eq. (1)).

 Notice the particular arrangement of double poles in Eq. (\ref{double}). 
 As we 
have indicated the irreducible part of the form factor  leaves certain ambiguity in whether to deform the poles into the upper or lower half plane. At present we have not been able to resolve this ambiguity.  The advantage of this particular
choice of poles is that  the integral can be straightforwardly evaluated  
(under the same assumption as before that the most important contribution
comes from small $\tp_i$ and $\Tt_i$). The result is 
\begin{eqnarray}
\la {\cal O}(x,t){\cal O}(0,0)\ra  &=&  
\la {\cal O}\ra^2\sum^\infty_{n=0} \frac{(-S(x,t))^n}{2^{2n}n!} \sum_{a_i}|f(a_1,\ldots,a_n)|^2 ,
\end{eqnarray}
where
\begin{equation}\label{S}
S(x,t) =  \frac{M}{\pi}\re^{-M\beta}\int d\tp \re^{-\frac{M\beta }{2}\theta^2}|x-t\tp|. \label{defS}
\end{equation}

We now turn to the computation of the residues, $f(a_1,\ldots,a_n)$, of the poles.
To simplify the discussion, we will consider only the case where the excitation spectrum 
of the theory constitutes a doublet: $a = \s = \pm 1$. 
This situation includes many important models, in particular the sine-Gordon model. 
In that case the semi-locality index of operator ${\cal O}$ with a soliton/anti-soliton equals $\pm \phi$.  We have, for instance,  
\[
f(++) = (1-\re^{\ri\phi})^2, ~~ f(+-) = 2(1 - \re^{-\ri\phi}), ~~ f(-+) = 2(1 - \re^{\ri\phi}), ~~ f(--) = (1-\re^{-\ri\phi})^2.
\]
\noindent In this case Eq. (\ref{solution}) reduces to
\bea
\sum_{\s_i = \pm 1}|f(\s_1,...\s_n)|^2 = 2^{n+2}\sin^2(\frac{\phi}{2})\left\{C_{2n-2}^{n-1} + 2\sin^2(\frac{\phi}{2})\sum_{k > j}C_{n-1}^{k-1}C_{n-1}^{j-1}(-\cos\phi)^{k-j-1}\right\}.
\eea
To evaluate the sum, $k<j$, we rewrite it as follows:
\begin{eqnarray}
2\sum_{k > j}C_{n-1}^{k-1}C_{n-1}^{j-1}(-\cos\phi)^{k-j-1} 
&=& \frac{1}{\cos(\phi)}C_{2n-2}^{n-1}-\frac{1}{\cos\phi} 
\sum^n_{k,j}C_{n-1}^{k-1}C_{n-1}^{j-1}(-\cos\phi)^{|k-j|};\cr\cr
&\equiv& \frac{1}{\cos(\phi)}C_{2n-2}^{n-1} - \Gamma .
\end{eqnarray}
For $\cos(\phi) > 0$, we use the identities
\begin{eqnarray}
\cos (\phi )^{|k-j|} &=& \frac{1}{\pi}\int^\infty_{-\infty} \rd\omega \frac{\delta}{\omega^2+\delta^2}\re^{i\omega (k-j)};\cr\cr
\sum^\infty_{m=-\infty} \frac{\delta}{4(\omega+m\pi)^2 + \delta^2} &=& 
\frac{1}{8}\frac{\sin^2(\phi)}{\sin^2(\omega)\cos(\phi)+\sin^4(\phi/2)},
\end{eqnarray}
where $\delta = - \ln\cos\phi$, to rewrite $\Gamma$ as
\begin{eqnarray}
\Gamma = \frac{2^{2n-4}}{\pi\cos(\phi)}\int^\pi_0\rd\omega \sin^{2n-2}(\omega)\frac{\sin^2(\phi)}{\sin^2(\omega)\cos(\phi)+\sin^4(\phi/2)}.
\end{eqnarray}
If we finally use the identity to write the combinatorial factor, $C_{2n-2}^{n-1}$, in a similar integral form
\begin{eqnarray}
C_{2n-2}^{n-1} = \frac{2^{2n-2}}{\pi} \int^\pi_0 \rd\omega (\cos(\omega/2))^{2n-2},
\end{eqnarray}
we can, using Eq. (\ref{S}), represent the correlator as 
\begin{equation}\label{asymp}
\la {\cal O}(x,t){\cal O}(0,0)\ra = \frac{1}{2\pi} \int^{\pi/2}_0 \rd\omega \exp\big[-2S(x,t)\sin^2(\omega)\big]
\frac{\sin^2(\phi)}{\sin^2(\omega)\cos(\phi)+\sin^4(\phi/2)}. \label{S1}
\end{equation}
It turns out that this relation holds even if
$\cos(\phi) < 0$.  To establish this we must use instead the identity
\begin{eqnarray}
\sum^\infty_{m=-\infty} \frac{\delta}{4(\omega+m\pi)^2 + \delta^2} &=& 
\frac{1}{8}\frac{\sin^2(\phi)}{\cos^2(\omega)\cos(\phi)+\sin^4(\phi/2)}.
\end{eqnarray}
where now $\delta = -\ln |\cos(\phi )|$.  We point out that as $\phi \rightarrow \pi$, the expression
for the asymptotics of the correlator (Eq. \ref{asymp}) reproduces the result for the Ising universality class.

  At $x=0$ Eq.(\ref{S1})  coincides with the correlation function obtained from the recent semi-classical analysis in Refs.\cite{kedarsubir,zarand}.  Both these papers explicitely assume that the two-particle S-matrix is equal to the permutation operator \cite{note1}. At finite $x$ however, the semiclassical  formulae are generically at variance with ours.  In particular, at $t=0$ 
   the semi-classical analysis rightly yields exponentially decaying  correlation functions. This  discrepancy presumably originates from a different  arrangement of poles in Eq.(\ref{double}). We hope to resolve this problem in a subsequent publication. As far as this paper is concerned, we will restrict the discussion to $x=0$ correlation functions.

\section{Low temperature charge dynamics in the sine-Gordon model}

To provide an example of a physical problem which can be solved by the method developed
in this paper, we consider the problem of charge counting statistics in the sine-Gordon model. 
The Lagrangian corresponding to this theory is 
\bea
{\cal L} = \frac{1}{16\pi}(\p_{\mu}\Phi)^2 - m^2\cos(\gamma\Phi) .
\eea
The cosine term is relevant at $\gamma < 1$. The conserved soliton charge is defined as 
\bea
Q = \frac{\gamma}{2\pi}\int \rd x\p_x\Phi .
\eea
Therefore $Q(x,t) = (\gamma/2\pi)[\Phi(x,t) - \Phi(x,0)]$ is the charge that has
passed through a point $x$ during time 
$t$. One can calculate the probability distribution 
function of $Q(x,t)$ using the results 
for the correlation function of  operators $\exp[\ri\eta\Phi]$. 
In this case $\varphi_{\pm} = \pm (\eta/\gamma)$. 

The sine-Gordon model is a good example to study as it encompasses both universality classes
as $\gamma$ is varied.  
At the
so-called reflectionless points $\gamma^2_n = 1/(n+1)$, $n>1$, 
the sine-Gordon model model has a diagonal S matrix. 
At these points the model belongs to the same universality class as the Ising model. 
Away from these values of $\gamma$, the soliton-soliton scattering is not diagonal and the model falls into the 
second universality class (although for operators $e^{\pm i\Phi\gamma/2}$ where the semi-locality index is
$\pm1/2$ 
we revert to the Ising universality class).  In this case the physics is governed by the second universality class
provided we are at a sufficiently low temperature.  
The rapidity scale governing this crossover is
$$
\theta_0 \approx \frac{\nu}{2}\cosh^{-1}\big(\cos(\frac{2\pi}{\nu})+2|\sin(\frac{\pi}{\nu})|\big),
$$
where $\nu = \gamma^2/(1-\gamma^2)$.  If $\gamma$ is close to one of the reflectionless
points, i.e. $\gamma^2 = 1/(n+1) + \delta$, the crossover rapidity simplifies
to
$$
\theta_0 \sim \frac{n+1}{n}\sqrt{\pi\delta}.
$$
The corresponding temperature
at which the crossover 
occurs is $T^* \sim \pi M \frac{(n+1)^2}{2n^2}\delta$. 

We now consider the counting statistics
present in the sine-Gordon model when it falls into each of the two universality classes.

\subsection{Counting statistics at the reflectionless points}

At $\gamma^2 = 1/(n+1)$ we are in the Ising-like universality class.
The correlation function then takes the form, as follows
from Eq. (\ref{corr}),
\bea
\la \re^{\ri\eta\Phi(x,t)}\re^{-\ri\eta\Phi(0,0)}\ra = 
C^2(\gamma,\eta)\exp\left[-2\sin^2(\pi\eta/\gamma)
\int \frac{\rd p}{\pi}\re^{-\epsilon(p)\beta}|x - t\frac{\p\epsilon(p)}{\p p}|\right]\label{sineG},
\eea
where $\epsilon (p)$ is the soliton energy and 
the factor $C(\gamma,\eta)$ is equal to 
the vacuum expectation value of $\exp[\ri\eta\Phi]$. 
This formula remains valid even for $\gamma^2 \neq (n+1)^{-1}$ provided  $\eta = \gamma/2$. The quantity $C(\gamma,\eta)$ was calculated in Ref. \cite{lukyanov}:
\bea
&&C(\gamma, \eta) = \left[\frac{a M\Gamma(1/(2-2\gamma^2))\Gamma((2-3\gamma^2)/(2-2\gamma^2))}
{4\sqrt{\pi}}\right]^{2\eta^2}\times\nonumber\\
&&\exp\left\{\int_0^{\infty}\frac{\rd y}{y}
\left[\frac{\sinh^2(2\eta\gamma y)}{2\sinh(\gamma^2 y)\sinh y\cosh[(1 - \gamma^2)y]} 
- 2\eta^2\re^{-2y}\right]\right\} .\label{factor}
\eea
Here we have assumed that the correlator, $\la \re^{\ri\eta\Phi(x,t)}\re^{-\ri\eta\Phi(0,0)}\ra$, 
satisfies the short distance normalization, 
$$
\la \re^{\ri\eta\Phi(x,t)}\re^{-\ri\eta\Phi(0,0)}\ra \sim \frac{4a^{\eta^2}}{|x-y|^{4\eta^2}},
$$
where $a^{-1}$ is the effective cutoff in the theory.
$M \sim m^{1/(1 - \gamma^2)}$ is the soliton mass. 
In the field theory limit, $M a \ll 1$, and so
the $\eta$ dependence in Eq. (\ref{factor}) is dominated by the term $(Ma)^{2\eta^2}$. 
Thus the magnitude of the propagator has a strong maximum at small $\eta$. 

We now examine the generating functional $\la \exp[\ri\omega Q(x,t)]\ra$.
This is given by 
\begin{eqnarray}
\la \re^{\ri\omega Q(x,t)}\ra &=& 
\la \re^{\ri\frac{w\gamma}{2\pi}\Phi(x,t)}\re^{\ri\frac{w\gamma}{2\pi}\Phi(x,0)} \ra \cr\cr
&=& C(\gamma, \frac{w\gamma}{2\pi})^2 \exp\bigg[-2\sin^2(\frac{\omega}{2})\int\frac{\rd p}{\pi}
\re^{-\beta\epsilon(p)}|t\partial_p\epsilon (p)|\bigg].
\end{eqnarray}
In the limit of large time the above integral finds its predominant contribution
at small $\omega$.  We then can write
\begin{eqnarray}\label{omega}
\la \re^{\ri\omega Q(x,t)}\ra &=& (\alpha aM)^{(\frac{\omega\gamma}{\pi})^2}
\exp\bigg[-\omega^2|t|/\tau_0\bigg],
\end{eqnarray}
where we have written $C(\gamma , \frac{w\gamma}{2\pi})^2 
\sim (\alpha a M)^{(\frac{\omega\gamma}{\pi})^2}$ for
some constant $\alpha$ and
where $1/\tau_0 = (T/\pi)\re^{- M/T}$.

The  Fourier transform of Eq. (\ref{omega}) then 
yields a Gaussian distribution function for the charge,
\begin{eqnarray}
P(\tilde{Q}) \equiv \la \delta(\tilde{Q} - Q)\ra
&=& \int \rd\omega \re^{-\ri\omega \tilde Q}\la \re^{\ri \omega Q(x,t)}\ra \cr\cr
&\sim & \s^{-1/2}\exp\left[- \tilde{Q}^2/4\s\right],
\end{eqnarray}
where
$\s = (t/\tau_0) + \frac{\gamma}{\pi^2}\ln[1/(\alpha aM)]$.

\subsection{Charge counting statistics at $\gamma^2 \neq 1/(n+1)$}
Here we consider the sine-Gordon model when it falls into the second universality class.
As we have already mentioned, at $\gamma^2 \neq 1/(n+1)$ 
there is a crossover temperature $T^*$ 
below which one can replace the S-matrix of slow solitons 
by the permutation operator.  
Then we may use Eq. (\ref{S1}), finding that
\begin{equation}
\la \re^{\ri\omega Q(x,t)}\ra = C^2(\gamma,\frac{w\gamma}{2\pi})\frac{1}{2\pi}
\int^\infty_0 \rd\omega \exp[-4\omega^2t/\tau_0]\frac{\phi^2}{\omega^2 + \phi^4/16}.
\end{equation}
The Fourier transform of this expression, after some manipulation, can be written as
\begin{equation}
P(\tilde{Q}) = \pi\bar{t}^{1/2} \int^\infty_{-\infty} \rd y 
\frac{\re^{-\frac{\tilde{Q}^2}{4(A+|y|)} - y^2/\bar{t}}}{\sqrt{A+|y|}},
\end{equation}
where $\bar{t} = t/\tau_0$ and $A = \frac{\gamma^2}{\pi^2}\ln (a^{-1}/(\alpha M))$.
At small times, the integral sees its main contribution at small $y$, thus allowing
us to write
\begin{equation}
P(\tilde{Q})_{\bar{t} \ll 1} \approx \frac{\pi^{3/2}}{\sqrt{2A}}\exp\bigg[-\frac{\tilde{Q}^2}{4A}\left(1-\frac{\tilde{Q}^2\bar{t}}{32A^3}\right)\bigg].
\end{equation}
In this case the distribution is that of a static Gaussian in $\tilde{Q}$ with
small time dependent corrections.
At large times we can evaluate this integral in the saddle point approximation which yields 
\begin{equation}
P(Q(t))_{\bar{t}\gg 1} \sim \bigg(\frac{1}{\bar{t}\tilde{Q}^2}\bigg)^{1/6}
\exp\bigg[-\frac{3}{2^{5/3}}\bigg(\frac{\tilde{Q}}{\bar{t}^{1/4}}\bigg)^{4/3}\bigg].
\end{equation}
The reader may notice that quantum fluctuations 
play a significant role in modifying the Gaussian distribution of
the counting statistics seen in the simpler Ising-like reflectionless case.

\section{Conclusions}
 
In this paper 
we have found universal expressions for the asymptotics of correlation 
functions of order parameter operators. 
These operators have $T=0$ nonzero vacuum expectation values 
thus breaking some discrete symmetry present in the theory. 
We restricted our consideration to theories with well defined asymptotic particle states 
(this excludes such models as the super sine-Gordon or the O(2N+1) Gross-Neveu model). 
Among such models, we have found two characteristic types of order parameter dynamics. 
One type appears in the models with either diagonal S-matrices (these include, among other models, 
the quantum Ising model and the sine-Gordon model 
at the reflectionless points) or for operators of non-diagonal theories with trivial semi-locality factors. 
These dynamics essentially coincide with the semi-classical 
dynamics described in \cite{subir}. 
Another type of dynamics is described by Eq. (\ref{S1}) and appears in the models 
where the zero momentum S-matrix corresponds to pure reflection and the order parameter has a non-trivial
semi-locality factor. 
 Here the asymptotics of correlation functions,  as determined by the form-factors,  coincide with the  
semi-classical expressions found in 
Refs. \cite{kedarsubir} and \cite{zarand} only at $x =0$.  We believe the origin of this difference is in the particular choice of the formfactor regularization. We hope to settle this problem in a subsequent publication.

\section{Acknowledgments}

This research was supported by DARPA under the QuIST Program, by the 
Institute for Strongly Correlated and Complex Systems at BNL (BLA), and by 
the DOE under contract number DE-AC02 -98 CH 10886 (RMK, AMT).  
All the authors are grateful to the International Centre for Theoretical
Physics in Trieste, Italy, for its hospitality during which this project was
completed.
AMT is grateful to
Princeton University for hospitality and to S. A. Lukyanov for 
attracting his attention to Ref. \cite{kiev}. 
We are also grateful to H. Babujian, M. J. Bhaseen, K. Damle, F. H. L. Essler,  V. E. Korepin, S. Sachdev, F. A. Smirnov, G. Zarand, and especially to A. Shytov and V. A. Fateev
for fruitful discussions and interest in the work.

\end{document}